\documentclass[twocolumn, aps,prl,showpacs,superscriptaddress]{revtex4}
\usepackage[dvips]{graphicx}
\usepackage{indentfirst}
\usepackage{bm}
\usepackage{enumerate}
\usepackage{amsmath}
\usepackage{amssymb}
\usepackage{txfonts}
\usepackage{graphicx}
\usepackage{dcolumn}

\begin{document}

\oddsidemargin=-10mm

\title{Efficient excitation of  a symmetric collective atomic state with a single-photon through dipole blockade}
\author{Fang-Yu Hong}
\affiliation{Department of Physics, Center for Optoelectronics Materials and Devices, Zhejiang Sci-Tech University, Xiasha College Park, Hangzhou, Zhejiang 310018, CHINA}
\author{Shi-Jie Xiong}
\affiliation{National Laboratory of Solid State Microstructures and
Department of Physics, Nanjing University, Nanjing 210093, China}
\author{W.H. Tang}
\affiliation{Department of Physics, Center for Optoelectronics Materials and Devices, Zhejiang Sci-Tech University, Xiasha College Park, Hangzhou, Zhejiang 310018, CHINA}
\date{\today}
\begin{abstract}
 In the famous quantum communication scheme developed by Duan {\it et al.}[L.M. Duan, M.D. Lukin, J.I. Cirac, and P. Zoller, Nature (London) {\bf 414} 413 (2001)], the probability of successful generating  a symmetric collective atomic state with a single-photon emitted have to be far smaller than 1 to obtain an acceptable entangled state.  Because of strong dipole-dipole interaction between two Rydberg atoms, more than one simultaneous excitations  in  an atomic ensembles are greatly suppressed, which makes it possible to excite a mesoscopic cold atomic ensemble into  a singly-excited  symmetric collective  state accompanied by a signal photon  with near unity success probability, at the same higher-order excitations can be significantly inhibited.

\end{abstract}

%\keywords{collective enhancement, dipole blockade, atomic ensemble}

\maketitle
\section {I. INTRODUCTION}

Quantum information science holds promise for many fascinating potential applications such as the factorization of large numbers\cite{shor}, secure communication\cite{aeke}, and spectroscopic techniques with enhanced sensitivity \cite{dumk}  . As a promising candidate for quantum state engineering and for quantum information processing, recently atomic ensembles with a large number of identical atoms have  received much attention. Such a system can been used, e.g.,  for generation of significant spin squeezing \cite{aknb,jhjs,aklm} and entanglement \cite{ldjc, akep, bjak, kchd, lmdu}, for storage of quantum light\cite{mlsy,mfmd,clzd,dpaf}, and for long-distance quantum communication \cite{ldml, sras,zccs}. As against quantum information schemes employing single particles, those based on atomic ensembles have two assets: first,  to manipulate atomic ensemble with laser  is easier than to control a single particle; second, the coupling between atoms and some light mode in atomic ensembles can be greatly enhanced by a factor of $\sqrt{N_a}$ for certain atomic level structure because of many-atom interference effects \cite{ldml, ldjcp}, where $ N_a$ denotes the atom number in the ensemble. Due to this coupling enhancement, the atom coupling to light does not require a high-finesse microcavity, which, in spite of recent experimental advances \cite{kjva, cjho}, remains  a demanding technology.

In the  Duan-Lukin-Cirac-Zoller (DLCZ) protocol for long-distance quantum communication \cite{ldml}, entanglement in the elementary links is created by
recording a single photon produced indistinguishably by one of two atomic ensembles. The probability
$p$ of successful generating one excitation in two ensembles is related to the
fidelity of the entanglement, leading to the the condition $p\ll 1$
to guaranty an acceptable entanglement quality. But when
$p\rightarrow0$, some experimental  imperfections such as stray
light scattering and detector dark counts will contaminate the
entangled state increasingly \cite{zyyc}, and subsequent processes
including quantum swap and quantum communication become more
challenging for finite coherent time of quantum memory \cite{kchd}. To solve this problem, protocols based on single photon sources \cite{kchd, fhsx}  were suggested.

 For the generation of a single collective  excitation  in the atomic ensemble with high success probability $p$ at the same time with high fidelity,  we describe a method based on the strong dipole-dipole interaction between the atoms in the  Rydberg state. When an atom is in the Rydberg state, the transition of another atom to this Rydberg level  is inhibited due to the level shift arising from the strong dipole-dipole interaction between Rydberg atoms, which is the so-called dipole block \cite{mlmf, euta,agym}. This blockade can be used to generate entanglement between two atoms and  therefor  to accomplish two-bit quantum gates in individual-atom systems \cite{djjc}. Recently Brion {\it et al.} propose to encode quantum information on the collective state of multilevel atomic ensembles and implement one and two-bit gates through collective internal state transitions in the presence of the dipole blockade \cite{ebkm}. This blockade can also be employed for a controlled generation of collective atomic state, for nonclassical photonic states, and for a scalable quantum logic gates \cite{mlmf}. In this paper, we will show that a single-excitation symmetric collective state of a mesoscopic  cold atomic ensemble accompanied by a signal photon can be created  with near unity success probability, while higher-order excitations in the ensemble can be significantly suppressed in the presence of the strong dipole blockade.

There have been many models describing the interaction between atomic ensembles and optical beams, such as the cavity-QED models \cite{aknb, mlsy, ldml}, one-dimensional light propagation models \cite{mfmd,ldjc}, and three-dimensional free-space perturbation theory \cite{ldjcp} which can describe most of the current experiments on the atomic ensembles\cite{aklm, bjak,clzd, dpaf}. Here,  we  describe the interaction of light and mesoscopic cold atomic ensembles with strong dipole-dipole interaction also in the three-dimensional free-space configuration but  without applying perturbation expansion.

\section{II. THE LIGHT-ATOMIC ENSEMBLE INTERACTION}
\begin{figure}
\includegraphics[width=0.7\columnwidth]{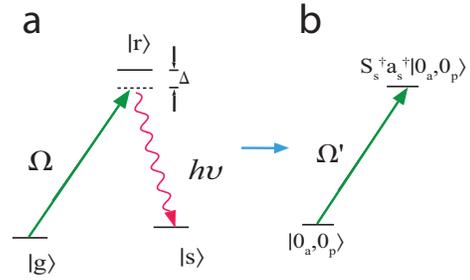}
\caption{\label{fig:1}(Color online) (a) The relevant atomic level structure with $|g\rangle$, a ground state, $|s\rangle$, a metastable state, and $|r\rangle$, a Rydberg state. A classical field with Rabi frequency $\Omega(t)$ is detuned the transition  $|g\rangle \rightarrow |r\rangle$ by $\delta$, while the transition  $|r\rangle \rightarrow |s\rangle$ is coupled to a signal light mode. (b) Under certain circumstance, the system consisting of  a cloud of $N_a$ atoms with the $\Lambda$ structure, a pumping laser, and spontaneous emission field is effectively equivalent to a two-level system under the action of the pumping laser with new Rabi frequency $\Omega'(t)$.}
\end{figure}

We consider a cold atomic ensemble of $N_a$ identical atoms contained in a volume $V$ with level structure shown in Fig.\ref{fig:1}a, ground state $|g\rangle$, metastable  state $|s\rangle$, and Rydberg state $|r\rangle$. Hereafter we will assume that the ground state and the Rydberg state are generated in a Doppler free configuration \cite{evgk}.  All atoms are initialized in the ground state $|g\rangle$. The $|g\rangle \rightarrow |r\rangle$   transition  is coupled by a classical Raman pumping laser $\varepsilon_p$ of central frequency $\omega_0$ and  Rabi frequency $\Omega({\bf r},t)\equiv e {\bf E}({\bf r},t)\cdot {\bf r_{rg}}/\hbar $, where $e$ is the electron charge, ${\bf E}({\bf r},t)$ is the slowly varying envelope of the pumping field amplitude, and $e{\bf r}_{rg}$ is the electric dipole moment of the corresponding atomic transition. The $|r\rangle \rightarrow |s\rangle$ transition is coupled to the spontaneous emission  field $\varepsilon_{s}$ which can be expanded into plane wave modes. In the interaction picture the Hamiltonian that models our system is thus
\begin{eqnarray}\label{eq1}
 H(t)&=&\Bigg\{ \sum^{N_a}_{i=1}\sigma_{sr}^i  \sum_{\bf k} g({\bf k})a^\dag_{{\bf k}} e^{-i[{\bf k}\cdot {\bf r}_i-(\omega_{{\bf k}}-\omega_0+\omega_{s}t)]}     \notag \\
  &+&  \sum^{N_a}_{i=1}\Omega({\bf r_i},t)\sigma_{rg}^i e^{ik_0z}+H.c. \Bigg\}+ \Delta\sum^{N_a}_{i=1} \sigma_{rr}^i,
\end{eqnarray}
where the pumping  field  $\varepsilon _p$ is assumed to  propagate  along the $z$ direction, $\Delta=\omega_r-\omega_0$ with $\omega_g=0$,  the pumping laser frequency detuning,  $\sigma_{\alpha \beta}^i=|\alpha\rangle_i\langle\beta|(\alpha,\beta=g,s,r)$, the transition operators of the $i$th atom, and the coupling factor  $g({\bf k})$ is dependent on the dipole moments of the corresponding transitions and  on the direction of the wave vector ${\bf k}$ of the spontaneous emission field  $\varepsilon_s$, which has a carrier frequency $\omega_0-\omega_s$ and frequency width determined approximately by the natural width $\Gamma$ of the Rydberg state $|r\rangle$. Thus the modes $a_{\bf k}$  with  $\omega_{\bf k}-(\omega_0-\omega_s)\gg \Gamma$ have scarcely any effect on  the system evolution. For the purpose specified in the the literatures \cite{ldml, ldjcp, zccs}, the spontaneous emission back to the ground state $|g\rangle$ is negligible.

On the condition that the detuning $\Delta$ substantially larger than the natural width $\Gamma$ of the level $|r\rangle$ and the frequency spreading of the classical field $\varepsilon_p$, we may use standard methods \cite{mors} to adiabatically eliminate the  level $|r\rangle$ from the system dynamics. The resulting Hamiltonian takes the form
\begin{eqnarray}\label{eq2}
 H(t)&=&-\Bigg\{ \sum^{N_a}_{i=1}\sigma_{sg}^i  \sum_{\bf k} \frac{\Omega({\bf r_i},t)g({\bf k})}{\Delta} a^\dag_{{\bf k}} e^{-i[\Delta{\bf k}\cdot {\bf r}_i-\Delta\omega_{{\bf k}}t]}     \notag \\
  &+&  H.c. \Bigg\}-\frac{1}{\Delta} \sum^{N_a}_{i=1} \sigma_{gg}^i|\Omega({\bf r_i},t)|^2,
\end{eqnarray}
where $\Delta \omega_{{\bf k}}=\omega_{{\bf k}}-(\omega_0-\omega_s)$ and $\Delta {\bf k}={\bf k}-k_0{\bf z}_0$ with the $z$ direction unit vector ${\bf z}_0$. Because the spontaneous emission field is greatly weaker than the pumping field,  the Stark shift of the  level $|s\rangle$ is far smaller than the other terms and has been omitted in the Eq.(\ref{eq2}). The last term in Eq.\eqref{eq2} corresponding to the Stark shift of the level $|g\rangle$ can  be removed  via a phase rotation of the basis $\{|g\rangle,|s\rangle\}$ which will map $\sigma_{sg}^i$ into $\sigma_{sg}^i e^{i \frac{1}{\Delta} \int ^t_0|\Omega({\bf r},\tau)|^2d\tau}$. Thus the Hamiltonian can be written in the form
\begin{eqnarray}\label{eq3}
 H(t)&=&- \Bigg\{\frac{1}{\Delta}\sum^{N_a}_{i=1}\sigma_{sg}^i\Omega({\bf r_i},t)e^{i \frac{1}{\Delta} \int ^t_0|\Omega({\bf r_i},\tau)|^2d\tau}       \notag \\
  &&\cdot\sum_{\bf k} g({\bf k}) a^\dag_{{\bf k}} e^{-i[\Delta{\bf k}\cdot {\bf r}_i-\Delta\omega_{{\bf k}}t]}+  H.c.\Bigg\} ,
\end{eqnarray}

Typically, the energy level $\omega_s$ is either around GHz or zero, which relies on whether the hyperfine level or the Zeeman sublevel  is chosen for the level $|s\rangle$ \cite{ldjcp}. The magnitude of the wavevector $k$ of the spontaneous emission field  in the summation $\sum_{\bf k}$ is about $kc=\omega_k\in[\omega_0-\omega_s-\upsilon/2,\omega_0-\omega_s+\upsilon/2] $, where $\upsilon$, the bandwidth of the field, is in the order of the natural width $\Gamma$ of the excited level $|r\rangle$ and can be smaller than 1GHz, $c$ is the speed of light in the vaccum. Thus we  may assume $\omega_k=\omega_0-\omega_s$. The length of the atomic ensemble $L$ is assumed to be smaller than $L\leq 10\mu $m,  then we have $(k-k_0)L\leq3.5\times 10^{-5}$. The amplitude $E_p$ of the pumping laser field can be adjusted to be  independent of the  coordinate ${\bf r_i}$ of the atoms in the ensemble, i.e. $\Omega({\bf r}_i,t)=\Omega(t)$. Thus, the Hamiltonian of the system can be rewritten in the form
\begin{eqnarray}\label{eq4}
 H(t)&=&-\left(\sqrt{N_a}\Omega'(t)S_s ^\dag a_s^\dag +H.c.\right)  \notag\\
 &&-\Bigg\{\frac{1}{\Delta}\sum^{N_a}_{i=1}\sigma_{sg}^i\Omega(t)e^{i \frac{1}{\Delta} \int ^t_0|\Omega({\bf r_i},\tau)|^2d\tau}       \notag \\
  &&\cdot\sum_{{\bf k}'} g({\bf k}) a^\dag_{{\bf k}} e^{-i[\Delta{\bf k}\cdot {\bf r}_i-\Delta\omega_{{\bf k}}t]}+  H.c.\Bigg\} ,
\end{eqnarray}
where
\begin{equation}\label{eq5}
    S_s=\frac{1}{\sqrt{N_a}}\sum_{i=1}^{N_a}|g\rangle_i\langle s|,
\end{equation}
\begin{equation}\label{eq6}
    \Omega'(t)=\frac{1}{\Delta}\Omega(t)e^{i \frac{1}{\Delta} \int ^t_0|\Omega(\tau)|^2d\tau}g(k{\bf z_0 }),
\end{equation}
$a_s$ is the annihilation operator of the signal field mode which is the spontaneous emission field mode propagating in the $z$ direction, and $\sum_{{\bf k}'}$ denotes that the summation of the wavevector ${\bf k}$ does not include the signal field mode.

We see that the Hamiltonian can be divided into two parts: the first part describes the coherent collective interaction between the  atomic ensemble and the signal field with an enhancing factor $\sqrt{N_a}$, the second part describes the incoherent interaction between individual atoms and the spontaneous emission field (not including the signal field). To contain about $N_a\sim 10^4$ cold alkali atoms within $V<(10 \mu {\text m})^3$ through magnetic or optical traps is within the reach of  current technologies \cite{mlmf, ebkm}. Note that such an atomic ensemble is still dilute enough to fulfill the condition $k/\sqrt[3]{\rho}\gtrsim1$, which ensures that there is no superradiance, where  $\rho$ is  the atomic number density.   According to the the literature \cite{ldml},  the signal-to-noise ratio $R_{sn}$ between the coherent interaction rate and the incoherent rate can be estimated as
 $R_{sn}\sim3\frac{\rho L}{k^2}$.
To estimate the magnitude of the incoherent interaction,  we assume $3\rho=10^3 \mu$m$^{-3}$, $L=8\mu$m, and $\lambda=0.5 \mu$m. Then we have $R_{sn}\sim 1.5\times 10^2 $. Note that the incoherent action only affects  the success probability and doesn't influence the fidelity of the obtained signal state \cite{ldml}.
Thus the incoherent individual atom-light interaction can be negligible compared with the enhanced coherent collective interaction, leading to a simpler Hamiltonian
\begin{equation}\label{eq7}
   H(t)=-\left(\sqrt{N_a}\Omega'(t)S_s ^\dag a_s^\dag +H.c.\right).
\end{equation}

This Hamiltonian describing a two-level system dynamics with the corresponding level $|0_a,0_p\rangle$ and $S^\dag_sa_s^\dag|0_a, 0_p\rangle$ (Fig. \ref{fig:1}b), where $|0_a\rangle$ denotes the ground state of the atomic ensemble $\otimes_i|g\rangle_i$ and $|0_p\rangle$, the vacuum state of the signal light mode. The single-quantum collective excitation $ S^\dag_sa_s^\dag|0_a, 0_p\rangle$ can now be generated by a  $\pi$ pulse, $\int^T \Omega'(\tau) d\tau=\pi$ .

\section{III. THE PROBABILITY OF HIGHER-ORDER EXCITATIONS}
Now we discuss the higher-order excitations in the atomic ensemble during the process of  generating single-quantum collective excitation. Because  of the strong dipole-dipole interaction, the energy level of Rydberg excited atoms with separations of several $\mu$m $|r\rangle$ is strongly  shifted. Thus the presence of one Rydberg atom is enough to inhibit the excitation of all other atoms in the atomic ensemble. Doubly Rydberg excited collective  states is defined as \cite {mlmf}
\begin{equation}\label{eq8}
   |S_r^2\rangle\equiv\sqrt{\frac{N_a}{2(N_a-1)}}S_r^{\dag 2}|0_a\rangle,
\end{equation}
where
\begin{equation}\label{eq9}
    S_r=\frac{1}{\sqrt{N_a}}\sum_{i=1}^{N_a}|g\rangle_i\langle r|.
\end{equation}
The probability $p_d$ of populating the doubly excited state $|S_r^2\rangle$ is smaller than $1\%$ for the situation where about $N_a\sim10^4$ cold alkali atoms is contained in $V<(10 \mu{ \text m})^3$ and $ T<100$ ns \cite{mlmf, mstw}. According to a recent experimental report \cite{euta}, the double-excitation probability  for two Rydberg atoms with  $n=90$ and separated by 10.2 $\mu$m is 0.02. Considering the probability $p_{1r}$ of exciting a  Rydberg state  of an individual atom in the ensemble is far smaller than that of generating a single excited collective state $|S_r\rangle$, the probability $p_{2r}$ of exciting two atoms into Rydberg states (not collective state)  is about $p_{2r}\sim p_{1r}^2$ \cite{ldml}, and thus can be absolutely  negligible. Further, considering that the Raman pumping laser is far detuned from the transition $|g\rangle\rightarrow|r\rangle$, the  probability $p_1$ of an atom populates  the level $|r\rangle$ is very low, and the probability of two atoms are excited in the Rydberg state are far more smaller than $p_1$. Thus the doubly-excited collective state is absolutely negligible,  if the duration $T$ of the pumping laser field is no longer than the life-span of the Rydberg state $|r\rangle$, which rules out the possibility with which two atoms in the ensemble are excited into the level $|r\rangle$ in sequence. Thus we can generate a single-excited collective state $S_s^\dag a_s^\dag|0_a,0_p\rangle$ with near unity success probability at the same time with near unity fidelity.

Because the Rydberg states, in the ideal limit, never doubly  populated, the scheme does not involve mechanical interaction between atoms and leaves the atomic internal state decoupled from atomic motion \cite{djjc}. Thus the temperature  only have to meet the condition that the atomic distribution should not vary substantially on the  duration of the  Raman pumping laser, which is the case for the temperature to be as high as a few mK \cite{mlmf}.

\section {IV. CONCLUSION}
 We  have present a method to describe the interaction between cold atomic ensembles and optical beams in the present of strong dipole blockade. The single-excited collective atomic state with a forward-scattered Stokes photon propagating along the same direction as the pumping laser does can be excited with a near unity success probability, at the same time, multi-excitation states can be significantly suppressed in the presence of the strong dipole blockade. With this advantage, this system can be used as an efficient on-demand single-photon source \cite{mstw,blmo} which would give access to arbitrary small absorptions \cite{epjc}, improve the quality and the generation rate of  random numbers \cite{jrpo}, and play a fundamental role in quantum information processing, from Bell's inequality test \cite{jpdb}, quantum teleportation \cite{dbjp}, linear-optics quantum computing \cite{ekrl,traw},  and quantum communication \cite{kchd}, to quantum cryptography \cite{gbnl}. Particularly, this scheme may open up a new possibility for scalable long-distance quantum communication.

\section {ACKNOWLEDGMENGS}
 This work was supported by the National Nature Science Foundation of China (Grant No. 10874071, 50672088 and 60571029) and  by Scientific Research Fund of Zhejiang Provincial Education Department (Grant No. Y200909693).

\end{document}